\documentclass[aps]{revtex4}
\usepackage{graphicx} 
\newcommand\vexp[1]{\big\langle\,#1\,\big\rangle}

\newcommand\op[1]{{{\bf #1}}} 
\newcommand\opc[1]{{\cal #1}}

\def \ee {\end{equation}}
\def \be {\begin{equation}}
\def \H {{\cal{H}}}

\newcommand \Tr[1]{{\tt Tr}\{ {#1}\}}
\begin{document}

\title{Evidence of several dipolar quasi-invariants in Liquid Crystals}
\author{C.J. Bonin$^{a,b}$, C.E. Gonz\'alez$^{a,c}$, H.H. Segnorile$^{c,a}$ and R. C. Zamar$^{a,c}$}        
\date{\today}
\affiliation {$^{a}$Facultad de Matem\'atica, Astronom\'{\i}a y F\'{\i}sica,
Universidad Nacional de C\'ordoba (FaMAF)
 M.Allende y H. de la Torre - Ciudad Universitaria, X5016LAE - C\'ordoba, Argentina \\
$^{b}$ Instituto de desarrollo tecnológico para la industria - CONICET - Santa Fe, Argentina \\
$^{c}$ Instituto de Física Enrique Gaviola - CONICET - C\'ordoba, Argentina.
  }
\begin{abstract}
In a closed quantum system of $N$ coupled spins with magnetic quantum number $I$, there are about $(2I + 1)^N$ constants
of motion. However, the possibility of observing such quasi-invariant (QI) states in solid-like spin systems in Nuclear Magnetic Resonance (NMR) is not a strictly exact prediction. The aim of this work is to provide experimental evidence of several QI, in the proton NMR of small spin clusters, besides those already known Zeeman, and dipolar orders (strong and weak). We explore the spin states prepared with the Jeener-Broekaert pulse sequence by analyzing the time-domain signals yielded by this sequence as a function of the preparation times, in a variety of  dipolar networks. We observe that the signals can be explained with two dipolar QIs only within a range of short preparation times. At longer times the time-domain signals have an echo-like behaviour. We study their multiple quantum coherence content on a basis orthogonal to the z-basis and see that such states involve a significant number of correlated spins. Then we show that the whole preparation time-scale can only be reconstructed by assuming the occurrence of multiple QI which we isolate experimentally.  
\end{abstract}

\maketitle
\section{Introduction}\label{introduction}

Quasi-equilibrium states (QE) found in Nuclear Magnetic Resonance (NMR) of strongly interacting nuclear spin systems are quantum states  which do not evolve under the system Hamiltonian, thus they can be represented by a reduced spin density operator diagonal in blocks in the eigenbasis of the spin-environment Hamiltonian \cite{SegZam11,walls06}. These states only evolve due to spin-lattice relaxation, over a time scale much longer than the one of the build-up of the quasi-equilibrium. An experimental NMR procedure to prepare and detect {\em dipolar} QE 
states in high-field solid-state NMR is the Jeener Broekaert pulse sequence (JB) \cite{jeene67}. Briefly, this technique consists of the  phase-shifted radiofrequency (rf) pulses: $90_x - \tau - 45_y - t_e - 45_y - t$. The first pulse creates single quantum coherences in the spin system, which evolve in the rotating frame mainly under the dipole spin-spin Hamiltonian during  $\tau$. Along this period, multi-spin single-quantum coherences can develop, and  the second $45_y$ pulse transforms part of the coherences just created into multi-spin order  \cite{baum85,walls06,cho03}. Two important processes occur during the evolution period $t_e$, along different timescales: decoherence and relaxation.  Finally, the third pulse converts QE states into observable single quantum coherence.

JB experiments are frequently used to  transfer the Zeeman order into dipolar order, which, in a variety of samples is the only observable dipolar QE state. It is an experimental fact that for short preparation times $\tau \ll 1/\omega_D$ in the JB sequence, the dipolar signal observed after the third pulse is proportional to the time derivative of the FID signal \cite{FID},  which indicates that the prepared state is very similar to the secular (high field) dipolar energy. 
Two dipolar QE states, were instead observed both on hydrated salts \cite{eisen78,keller}, and on liquid crystals (LC) \cite{schmie82,interpar}. In the former, the occurrence of `intra-pair' and `inter-pair' QE states was associated with the distribution of the proton pairs of the water molecules in the lattice, conforming a system of weakly interacting spin pairs. 
The spin system of LCs is composed by the few proton spins within each molecule, which display a hierarchy of dipolar couplings. This kind of dipolar network is different from hydrated salts and a model of spin pairs does not strictly proceed, however, the dipolar Hamiltonian can in fact be partitioned into two mutually commuting and orthogonal parts, the `strong'  $\H_{\opc{S}}$ and the `weak' $\H_{\opc{W}}$ terms 
 \cite{ssnmr09}. Since these operators are chosen so that they commute with the total spin Hamiltonian, they are  nearly constants of the motion (for evolution times $t_e$ much greater than $1/\omega_D$ \cite{weitek82}). For this reason, they are called quasi-invariants (QI). %
 
In early works on dipolar order relaxation in LC's, the experiment was interpreted in terms of a single quasi-invariant, $\rho_{\cal S}$, of an ensemble of isolated representative strongly coupled spin pairs \cite{pintar74}; however, at that time, the dependence of relaxation times on $\tau$ was not understood \cite{Noack96}.
Later work on LC showed that the state $\rho_{\cal W}$ prepared by setting the preparation time to the one at which the $\rho_{\cal S}$ order vanishes,  also presents properties of quasi-equilibrium. In fact, it relaxes exponentially towards equilibrium, with a characteristic time which is also of the order $T_{1D}$ \cite{interpar}. Besides,  experiments  where the  multiple quantum coherences of the states prepared with the JB sequence are encoded in orthogonal, $X$ basis, carried out in nematic 5CB  \cite{bulj09}, confirmed that the tensor structure of the $\cal S$ state corresponds to two-spin dipolar order while the $\cal W$ state shows high order quantum coherences, all of which relax towards equilibrium with the same decay constant $T_{1\opc{W}}$. This indicates that for short preparation times the two-spin order prevails while higher order spin correlations are growing. This second QI, produces discernible signals coresponding to multi-spin order, created within a time window $\tau$ where the dipolar order vanishes, similarly to the hydrated crystal studied in reference \cite{keller}. 

Recently, it was shown that quantum decoherence can provide an efficient mechanism by which the spin system attains a diagonal state in the basis of the spin-environment Hamiltonian, over a time scale which is intermediate between those governed by its own interactions, and thermalization ruled by thermal fluctuations of the environment \cite{nosPRE11,SegZam11}. All these results support the assumption that in LC's, for times $t_e$  greater than the decoherence time scale,  the density operator can be written in the form \cite{ssnmr09}
\be  \rho_{D}= \op{1}-\beta_{Z}\H_{Z}-\beta_{\opc{S} }(\tau)\H_{\opc{S}}-\beta_{\opc{W}}(\tau)\H_{\opc{W}} \,. \label{rhoQE_SW}
\ee 
In fact, the experimental behaviour of the dipolar signal of 5CB can be described with Eq.(\ref{rhoQE_SW}) for preparation times  $\tau \leq 80 \mu$s \cite{interpar}. 
Relaxation brings the system to thermal equilibrium with the whole surrounding world over a much longer time scale, $ t_e\gg \tau$, through processes involving energy exchange between the spins and the lattice. The characteristic lifetime $T_{1D}$ of QE states (dipolar order relaxation time) is comparable to the common Zeeman relaxation time.

Theoretically, in a cluster of $N$ dipole interacting spins 1/2, as the protons of a typical LC molecule, there are at least $2^N$ constants of motion, the exact number depends on the degeneration of the dipolar Hamiltonian \cite{walls06}. That is, such a number of spin operators are needed to span the commutative space (or diagonal in blocks space) of the dipolar Hamiltonian.
The results obtained so far: the occurrence of two dipolar QI, are consistent with this idea, but accordingly, it should be possible to prepare new quasi-equilibrium states from the initial Zeeman order, which should be observable for longer preparation times in the JB experiment, as long as the JB signal is detectable. 

The aim of this work is to explore the spin states prepared for long preparation times in the JB experiment, looking for experimental evidence of multiple QIs that can be expected for dipole coupled spin clusters. The analysis is based on an exhaustive study of the time-domain signals in a variety of dipolar networks.

Providing insight on the physics nature of the quasi-invariants of a spin cluster can be useful both for applications as for basic research.
QE states are  relevant observables of the spin system, providing relaxation parameters useful to study molecular motion in LC mesophases through their dependence on temperature and magnetic field \cite{jcp98,jcp05,interpar}
They have been proposed as initial states for the excitation of multiple quantum coherences (MQ)
 in MQ-NMR \cite{emid80,doronin07,furman05} and used in spin counting experiments \cite{bulj09,cho03}. 
Besides, these states have been proposed as alternative to the Zeeman order in Magnetic Resonance Imaging \cite{matsui03}. 
Implementation of noiseless quantum memories and multi-spin quantum register relies on the possibility of manipulating multi-spin correlated entities which are unperturbed by decoherence processes \cite{mahesh06,krojanski06,lidar98,khitrin02}. 
From a basic viewpoint, the physics of systems with few degrees of freedom coupled to a quantum environment  attracts today's attention of a widespread community because of the potentiality for applications such as quantum devices and quantum information processing \cite{nielsen01,suter08,mahesh06} and, significantly, also because these systems are testbeds for studying fundamental aspects as irreversibility, equilibration and thermalization \cite{yukal12,polko11,SegZam11}.

 Section \ref{QI} contains the definition of the QI operators and their relation with the NMR signals in the JB experiment.   In Section \ref{Exp} we present an experimental survey of the manifestation of multiple quasiinvariants and propose a method to isolate them in samples with different geometries. The nature of the different QI is examined by means of relaxation experiments and by studyng the multiple quantum coherence content on the x-basis.

\section{Quasi invariants}\label{QI}
The spin Hamiltonian of resonant nuclei in ordered systems like solids and liquid crystals has an important contribution from the dipole-dipole coupling energy besides the Zeeman term,
\be \opc{H}_S= \opc{H}_Z +\opc{H}_D. 
\ee %. 
where the Zeeman energy in units of $\hbar$ is $\H_{Z}=-\omega_{o} {\bf I}_{z}$, with $\omega_{o}$ the Larmor frequency. In the rotating frame description \cite{MLevitt}, the time evolution of any spin state is mainly driven by the dipolar Hamiltonian and often only by its secular part (high field approximation),
\be \label{HD0}
\H_{D}^0= \sqrt{6}\sum_{i<j}D_{i j}{\bf T}_{20}^{i j} .
\ee
In liquid crystals the secular part of the dipolar coupling between nuclei $i$ and $j$ is 
$$%\be
D_{i j}\equiv \bigg\langle \frac{\mu_{o}\gamma^{2}\hbar}{4\pi}\left(\frac{1-3\cos^{2}\theta_{i j}}{2 r_{i j}^{3}}\right)\bigg\rangle
\label{dipcap}
$$ with $r_{ij}$  the distance between spins, $\theta_{i j}$  the angle between the internuclear vector and the magnetic field and the angle brackets stand for an average over internal and reorientational molecular motions. The residual dipolar coupling in LC only involves protons within a molecule, then, the sum of Eq.(\ref{HD0}) runs over  protons within a molecule.  
${\bf T}_{20}^{i j}$ is the zero component of a normalized irreducible spherical tensor of rank two, which in terms of the spin angular momentum  operators is \cite{abrag61,abragol}
$$%\begin{equation}
{\bf T}_{20}^{i j}=\frac{1}{\sqrt{6}}\left[ 2 I_{z}^{i}I_{z}^{j}-\frac{1}{2}\left(I_{+}^{i}I_{-}^{j}+I_{-}^{i}I_{+}^{j}\right) \right]\,.
$$%\end{equation}
Then, an arbitrary initial state $\rho_0$ of a closed system evolves, according to the Liouville equation, like
$$%\be \label{evrho}
\rho(t)=e^{-i \H_{D}^0t} \; \rho_0 \; e^{i \H_{D}^0t}.
$$

However, the interaction of a finite quantum system with the environment within the essentially adiabatic regime \cite{SegZam11} (or quasi-isolated regime \cite{yukal12}) attenuates the off-diagonal elements of the spin state (in the basis of the spin-lattice interaction Hamiltonian) while preserving the diagonal in blocks.  
We say that a spin system is in a state of quasi-equilibrium when its density operator in the rotating frame has a diagonal form as 
\be \rho_{qe}(\tau)= \frac{1}{\opc{N}} \left( \op{1}- \sum_k^{\sim 2^N} \beta_{k}(\tau) \opc{Q}_k \right) \label{rhoqe},
\ee
where $\opc{Q}_k$ are zero-trace operators which commute with the spin Hamiltonian in the rotating frame, $[\opc{Q}_i,\opc{H}_D^0]=0$ and satisfy the orthogonality relations
$$
\Tr{\opc{Q}_k\opc{Q}_{k'}}= \delta_{k,k'} \Tr{\opc{Q}_k\opc{Q}_k},
$$
the coefficients $\beta_k(\tau)$ are 
$$
\beta_k(\tau)= -\Tr{\opc{Q}_k \: {\cal N}\: \rho_{qe}(\tau,t)}/ \Tr{\opc{Q}_k^2},
$$
and $\cal{N}= \Tr{\op{1}}$ is the normalization factor.  

In the JB experiment, such a state of internal quasi-equilibrium is reached when a time after the second pulse $t_e$ longer than decoherence has elapsed. 
In this way, the coefficients $\beta_k$ can be interpreted as a generalization of the inverse spin temperatures \cite{abragol}. 
Since the QE states can be written in terms of the constants of motion $\{\opc{Q}_k\}$ these states only evolve in the long timescale of spin-lattice relaxation, so we call them {\em quasiinvariants} (QI) and their expectation values satisfy $\beta_k(\tau)=\vexp{\opc{Q}_k}/ \Tr{\opc{Q}_k^2}$.

The NMR signal detected in phase with the read pulse, compatible with Eq.(\ref{rhoqe}) is
\begin{equation} \begin{array}{rl}
S(t)= \vexp{{I}_y(t)} &=\Tr{U(t)\; P_r \;\rho_{qe}\; P_r^{\dag} \; U(t)^{\dag} \;  I_y } \\ \\
		      &= \sum_{i} \beta_i(\tau) \Tr{U(t) P_r \opc{Q}_i P_r^{\dag} U^{\dag}(t) I_y}\\ \\
		      &=\sum_i \beta_i(\tau) \opc{F}_i(t) ,\label{signal_qi}
 \end{array}
\end{equation} 
where $P_r$ represents the read pulse and \mbox{$U(t)= \exp\{-i\opc{H}_D^0 t\}$} is the evolution operator during the observation period. 
Within this view, coefficients $\beta_i(\tau)$ can be interpreted as the weight with which  operator $\opc{Q}_i$ contributes to the overall NMR signal for a given preparation time, and $\opc{F}_i(t) $ as the contribution of the $i$-th quasiinvariant to the dipolar signal. 
In fact, the Zeeman QI has no projection on the signals detected on the ``dipolar channel'' (channel $y$ in Eq.(\ref{signal_qi})) after the JB sequence, which is equivalent to considering $\beta_Z$=0.
  It is worth to mention that $S(t)$ is symmetric with respect to times $\tau$ and $t$. This property, of the JB sequence. In the case that the only QI is the total dipolar energy $\opc{H}_D^0$, as is the case of systems with regularly distributed nuclear spins, the dipolar signal of Eq.(\ref{signal_qi}) is proportional to the time derivative of the FID signal \cite{abragol}.  

The dipolar signals for $\tau < 80 \mu$s in both gypsum single crystal and nematic  4'-$pentyl$-4-biphenyl-carbonitrile (5CB)  were  accounted for in terms of two dipolar QI. The constants of motion in gypsum are the dipolar interaction energy of the spin pairs $\vexp{\opc{H}_{D\,M}^0}$ and  the interaction energy of each spin with all but its pair spin, $\vexp{\opc{H}_{D\,I}^0}$ \cite{dumont,eisen78,keller}, while the QI used in 5CB are the strong  $\vexp{\H_{\opc{S}}}$ and weak $\vexp{\H_{\opc{W}}}$ parts of the dipolar interaction \cite{interpar,ssnmr09,bulj09}. 

\section{Experimental} \label{Exp}

The interest herein is to explore the manifestation of multiple QIs after the Jeener-Broekaert rf pulse sequence. With this aim we analyze the JB signals as a function of the preparation time in different kinds of dipolar networks, with particular attention on long preparation times. 
Experiments at 20MHz were carried on in a Bruker Minispec mq20 and those at 60MHz in a homemade pulsed NMR spectrometer based on a Varian EM360 magnet \cite{segnotesis,bonintesis}.

\begin{figure}%[h*]
\includegraphics[width=10cm,height=7cm]{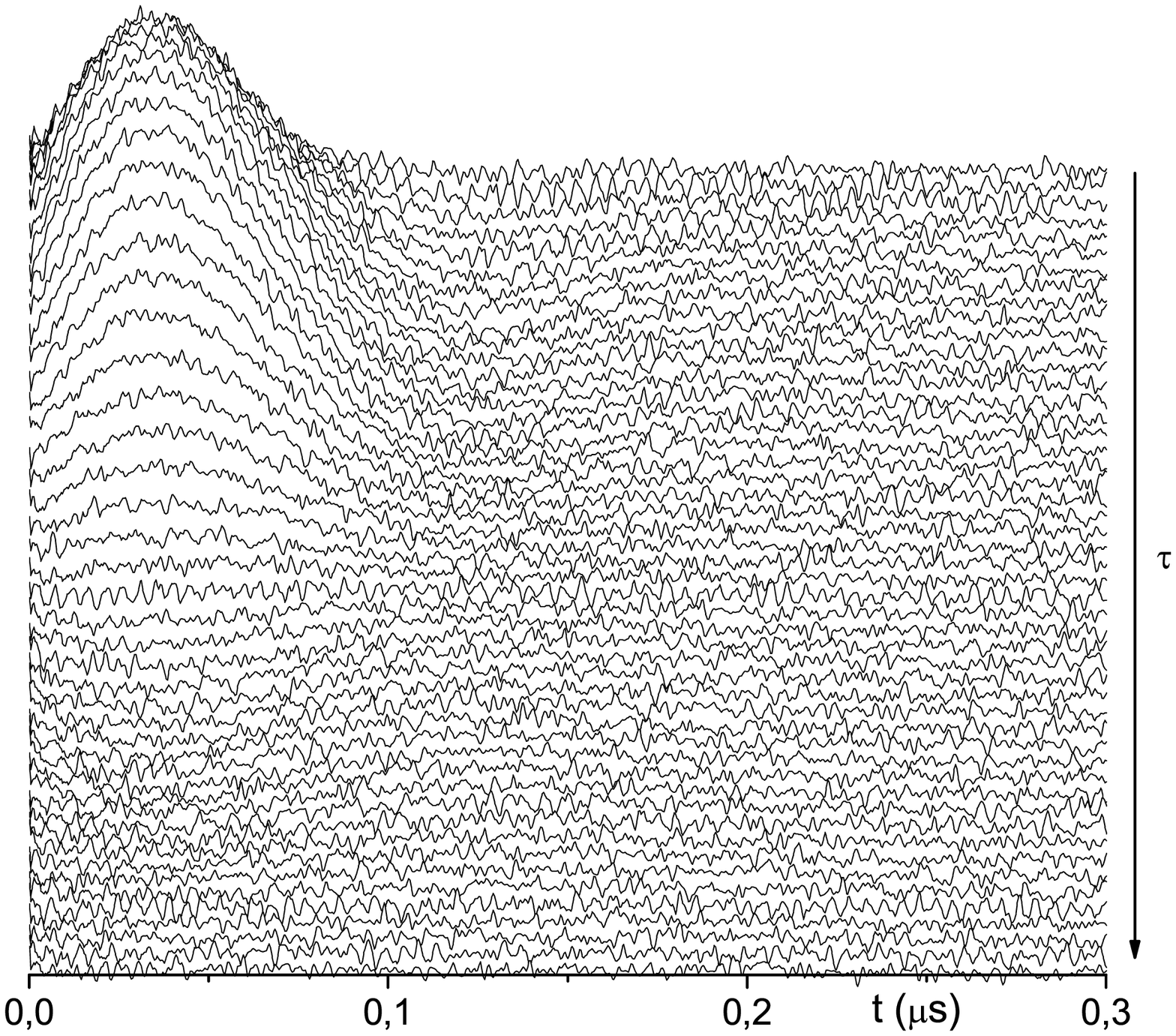}
\includegraphics[width=10cm,height=7cm]{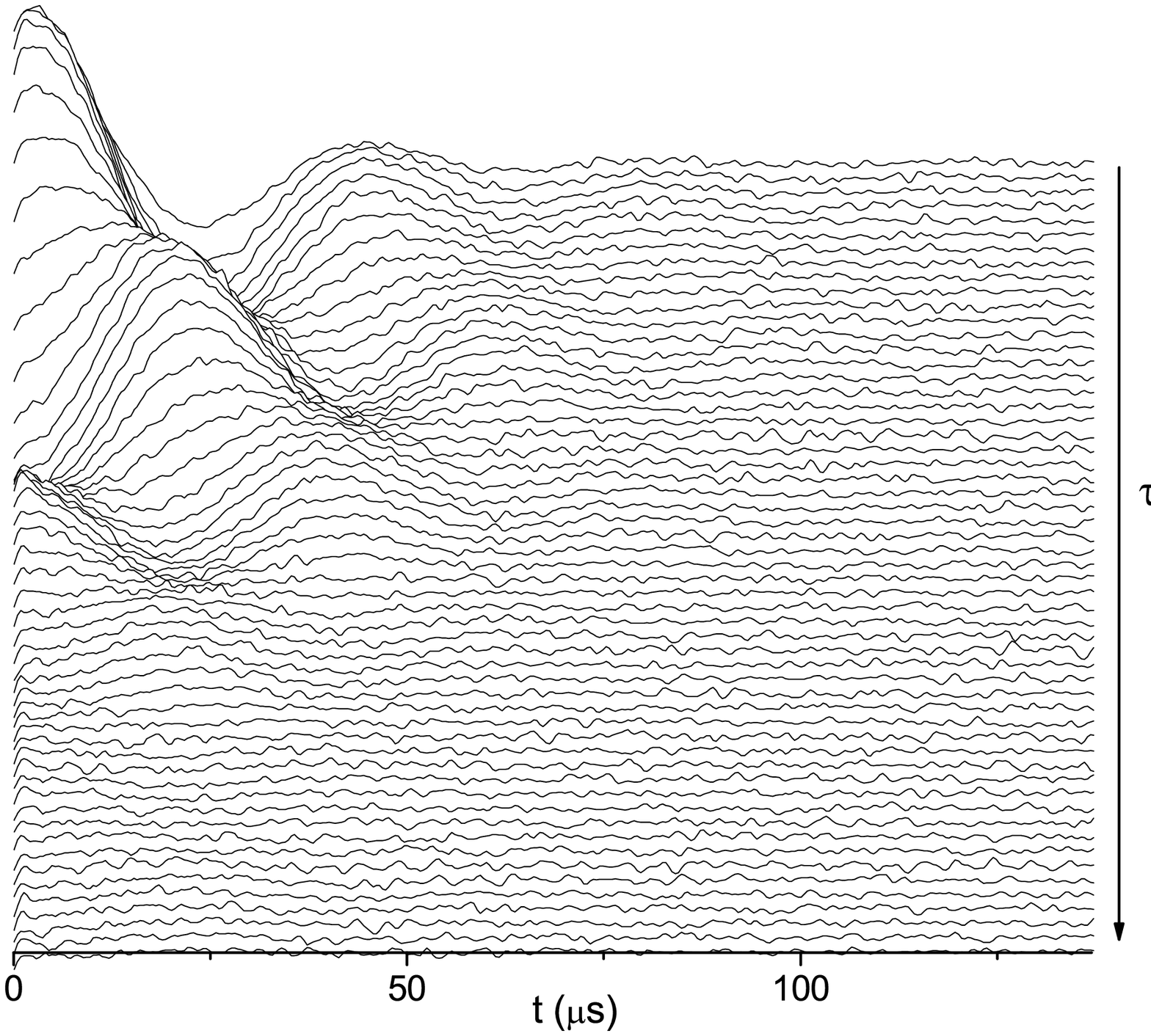}
\includegraphics[width=10cm,height=7cm]{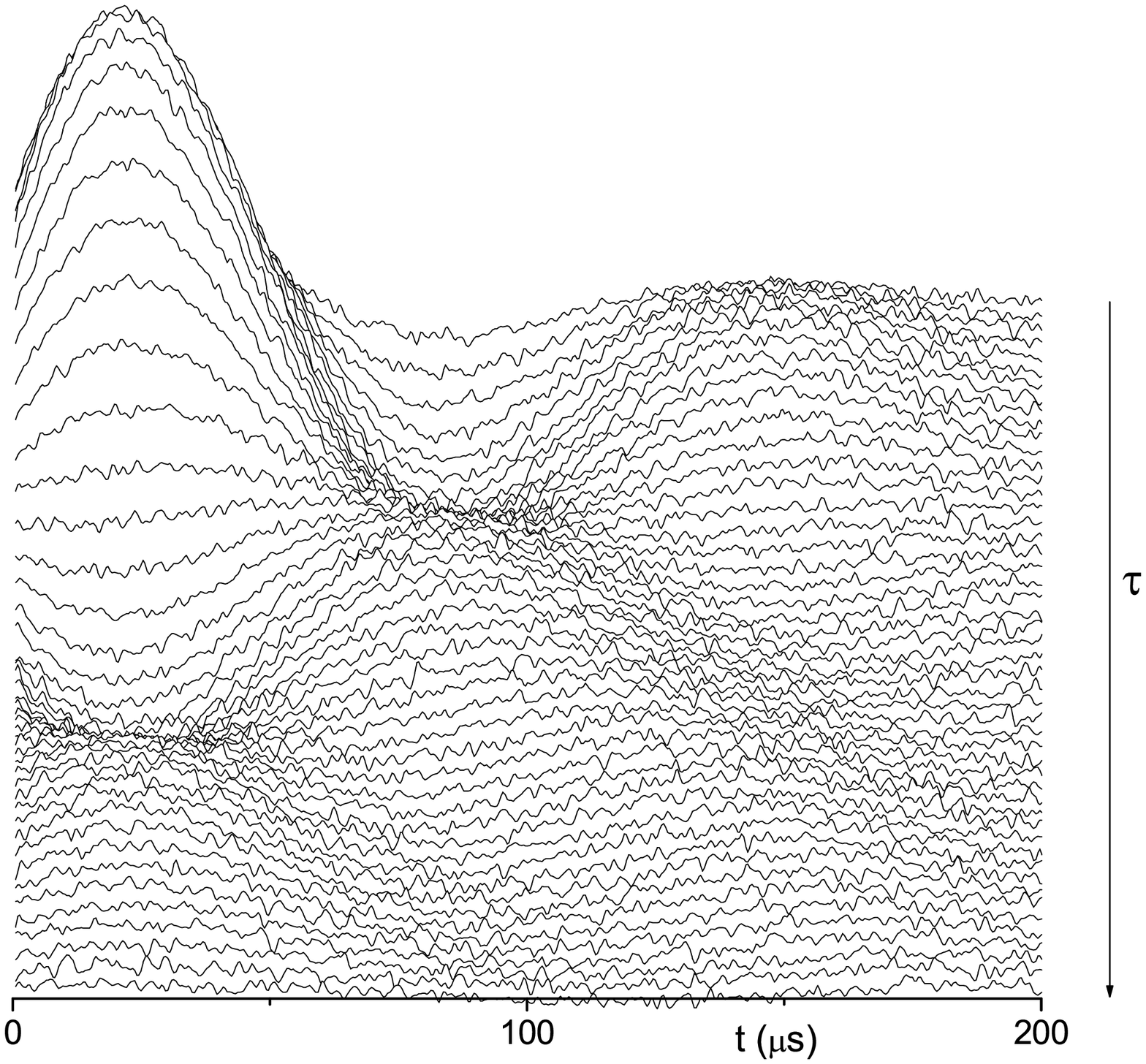}
\caption{Stackplots of the JB signals as a function of the preparation time $\tau$ in (a) Adamantane, (b) Gypsum and (c) MBBA liquid crystal.}
\label{dipolarland}
\end{figure}

As shown in Fig.\ref{dipolarland}, the response of the different compounds to variations of the preparation time is diverse, however they all share the characteristic of being symmetric with respect to $\tau$ and $t$. 
In powder adamantane (Fig.\ref{dipolarland}a), the signal is proportional to the time derivative of the FID and keeps this shape for every preparation time $\tau$, which is consistent with the 
occurrence of only one dipolar constant of motion: the dipolar energy, 
$\vexp{\opc{H}_D^0}$ \cite{jeene67,cho05}. 
%\begin{figure}[h*]
%\vspace{7 cm}
%\special{wmf:JB_adamant.wmf x=9cm y=7cm}
%\vspace{6 cm}
%\special{wmf:JB_gyp.wmf x=9cm y=7cm}
%%\vspace{-1 cm}
%\vspace{6 cm}
%\special{wmf:JB_MBBA.wmf x=9cm y=7.3cm}
%\caption{Stackplots of the JB signals as a function of the preparation time $\tau$ in (a) Adamantane, (b) Gypsum and (c) MBBA liquid crystal.}
%\label{dipolarland}
%\end{figure}

On the contrary, in the other compunds shown in Fig.\ref{dipolarland} [(b) Gypsum and (c) N-(4-Methoxybenzylidene)-4-butylaniline  (MBBA)], the dipolar signal shapes change drastically with $\tau$, being proportional to the time derivative of the FID only for short $\tau$ (more precisely, the proportionality holds while $\tau < t_0$, where $t_0$ is the time at which the time derivative of the FID first crosses through zero). 

However, a new behaviour of the dipolar signals arises clearly in LC for longer preparation times. The observed signals show an ``echo'' like behaviour, that is, they have a peaked shape whose maximum occurs at a time $t$ from the read pulse equal to the preparation time $\tau$. Such behaviour is not observed in gypsum. 

\begin{figure}%[h*]
\includegraphics[width=13cm,height=17cm]{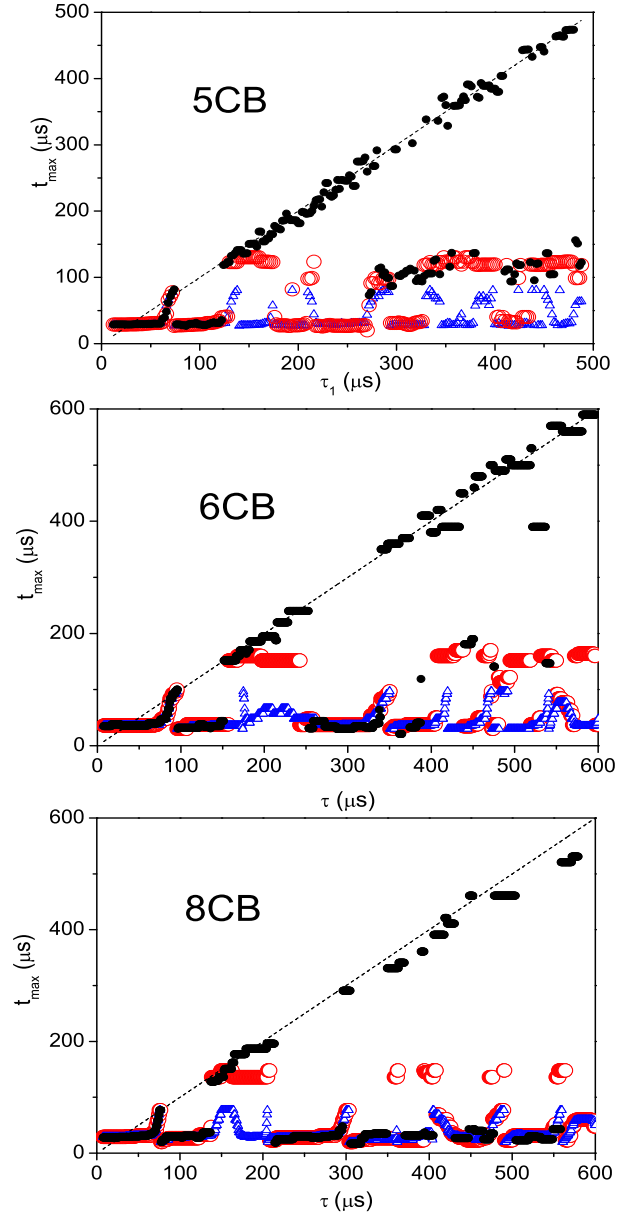}
\caption{Acquisition time $t_{max}$ at which the signals attain their maximum amplitude as a function of the preparation time $\tau$ in the nematic phase of  5CB (top), 6CB (middle) and 8CB (bottom). Solid circles represent the experimental data. Open triangles and open circles correspond to $t_{max}(\tau)$ simulated from Eq.(\ref{signal_qi}) with two and three quasiinvariants respectively. QIs selected as indicated in Table 1.}
\label{maxabs_1}
\end{figure}

\begin{figure}[h*]
\includegraphics[width=13cm,height=17cm]{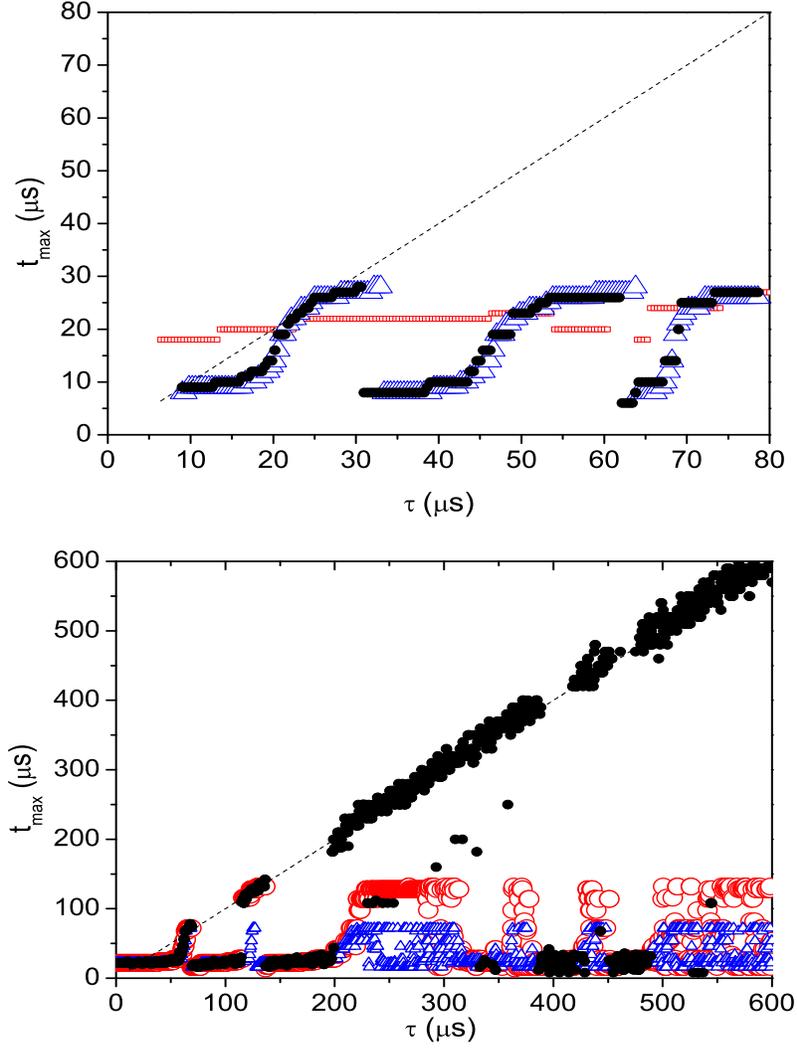}
\caption{Acquisition time $t_{max}$ at which the signals attain their maximum amplitude as a function of the preparation time $\tau$ in a gypsum single crystal (top) and in nematic MBBA (bottom). Solid circles represent the experimental data. Open triangles and open circles correspond to $t_{max}(\tau)$ simulated from Eq.(\ref{signal_qi}) with two and three quasiinvariants respectively. QIs selected as indicated in Table 1. Solid circles on the gypsum graph correspond an orientation $\vec B \parallel$ [010], while open squares to an orientation that makes the protons at the water molecules equivalent.  }
\label{maxabs_2}
\end{figure}

 In order to highlight this feature, Figs. \ref{maxabs_1} and \ref{maxabs_2} show the time $t_{max}$ at which the dipolar signal attains its maximum amplitude in the detection period, as a function of the preparation time $\tau$, in different compounds. Preparation times were varied within the range where the S/N ratio is greater than 1\% on each compound. It can be seen that the experimental $t_{max}(\tau)$ (solid circles in Fig. \ref{maxabs_1}) is very similar in all the cyanobiphenyl samples, 4'-$n$-4-biphenyl-carbonitrile where $n$ stands for pentyl (5CB), hexyl (6CB) and octyl (8CB). Data from nematic MBBA at the bottom of Fig. \ref{maxabs_2} behave just as the other nematic LC samples. This behaviour is drastically different in the gypsum single crystal (top of  Fig. \ref{maxabs_2}) since $t_{max}(\tau)$ is bounded in gypsum while grows linearly for long $\tau$ in all the nematic LC samples. As seen in this figure, the whole observable range on gypsum is rather smaller than that of the LCs, and the remarkable difference is that LCs develop the slope-one behaviour at preparation times near one fourth of their whole observable ranges while gypsum never does.  
It is also worth to note that the behaviour of $t_{max}(\tau)$ is also different in two orientations of the gypum single crystal with respect to the external magnetic field. The wavy curve of solid circles  in Fig. \ref{maxabs_2}(top) corresponds to the orentation $\vec B \parallel$ [010] while the open squares correspond to an orientation that makes the protons at the water molecules equivalent. 
%\begin{figure}[h*]
%\vspace{14 cm}
%\special{wmf:maxabs568CB.wmf x=10cm y=14cm}%maxabsqi3_5CB
%\hspace{-2 cm}
%%\vspace{-0.2 cm}
%\caption{Acquisition time $t_{max}$ at which the signals attain their maximum amplitude as a function of the preparation time $\tau$ in the nematic phase of  5CB (top), 6CB (middle) and 8CB (bottom). Solid circles represent the experimental data. Open triangles and open circles correspond to $t_{max}(\tau)$ simulated from Eq.(\ref{signal_qi}) with two and three quasiinvariants respectively. QIs selected as indicated in Table 1.}
%\label{maxabs_1}
%\end{figure} 
%
%\begin{figure}[h*]
%\vspace{10 cm}
%\special{wmf:maxabs_sele_MBBA_layout.wmf x=8cm y=10cm}%maxabsqi3_5CB
%\hspace{-2 cm}
%%\vspace{-0.2 cm}
%\caption{Acquisition time $t_{max}$ at which the signals attain their maximum amplitude as a function of the preparation time $\tau$ in a gypsum single crystal (top) and in nematic MBBA (bottom). Solid circles represent the experimental data. Open triangles and open circles correspond to $t_{max}(\tau)$ simulated from Eq.(\ref{signal_qi}) with two and three quasiinvariants respectively. QIs selected as indicated in Table 1. Solid circles on the gypsum graph correspond an orientation $\vec B \parallel$ [010], while open squares to an orientation that makes the protons at the water molecules equivalent.  }
%\label{maxabs_2}
%\end{figure} 

It has been already shown that two independent QI ($\opc{S}$ and $\opc{W}$) can be prepared in nematic LC and some hydrated salts \cite{dumont,interpar}, with the property that $\opc{W}$ emerges when the weight of $\opc{S}$ becomes zero. In fact, the set of dipolar signals in 5CB prepared with $\tau < 80 \mu$s were satisfactorily reproduced by using such  two QI in Eq.(\ref{signal_qi}) \cite{interpar}. However, this procedure does not describe the signal behaviour for $\tau > 80 \mu$s. 
Now, we propose a procedure to extract the quasiinvariants from the experimental data, under the sole assumption that the states prepared in the experiment are described by Eq.(\ref{rhoqe}).   It might be expected that new QI, if observable, will arise sequentially as the preparation time increases and that in a favourable condition, a subset of preparation times $\{\tau_i\}$ exists at which one QI (the $i$-th) prevails over the others. This amounts to proposing that it is experimentally possible to isolate a subset of QI. The validity of such working hypothesis will be tested along this work. 

The measured data sets are composed by the signals obtained for each of the $m$ chosen values of the preparation time $\tau=\tau_m$. Let us call $\opc{M}_{\tau_m}(t)$ to the acquired time domain signals after the JB sequence, which we arrange as the rows of the data array, and call 
$\opc{N}_{t}(\tau)$ to the columns of such array (pseudo-signals). Any cut $\opc{N}_{t_1}(\tau)$ at $t=t_1$ is found to be strictly symmetric with the signal  $\opc{M}_{\tau_1}(t)$ prepared with $\tau_1= t_1$. 

\begin{figure}[h*]
\includegraphics[width=13cm,height=9cm]{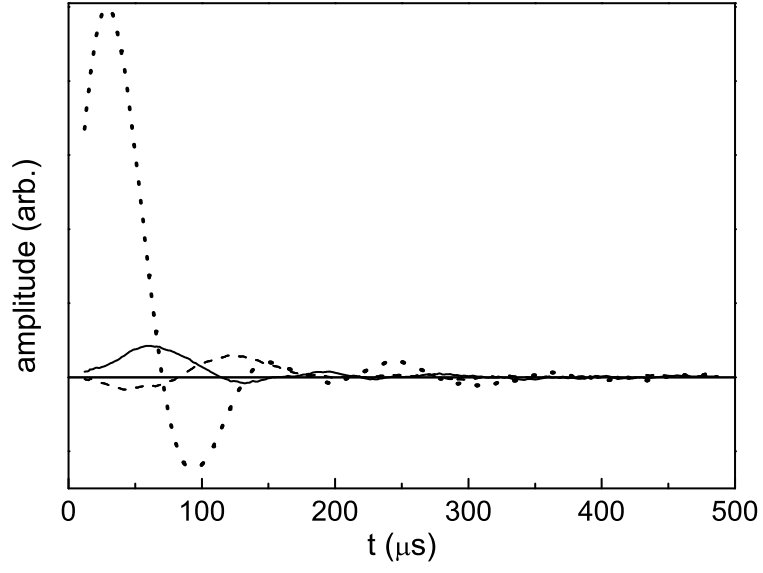}
\caption{Signals $\opc{F}_1$ (dotted), $\opc{F}_2$ (solid) and $\opc{F}_3$ (dashed) for 5CB at 302K.}
\label{funbase}
\end{figure}

 With the aim of deriving the functions $\opc{F}_i(t) $ from the experimental data we followed these steps: ($i$) Recognize the preparation time $\tau=\tau_1$ which makes $\opc{M}_{\tau_1}(t)$ proportional to the time derivative of the FID signal and has maximum amplitude. This signal attains its maximum at $t=t_1=\tau_1$. Based on former evidence that the state prepared with a time $\tau=\tau_1$ is a QI, we assume that the experimental function $\opc{M}_{\tau_1}(t)$ can be identified with the contribution of the first QI to the signal, $\opc{F}_1(t)$ (see dotted line in Fig. \ref{funbase}). ($ii$) Also select the experimental pseudo-signal $\opc{N}_{t_1}(\tau)$, which corresponds to a cut on the data array at an observation time $t=t_1$, and identify it with the ``weight'' $\beta_{1}(\tau)$ of the first QI. If the only dipolar QI is the dipolar energy, as is the case of adamantane, the dipolar signal of Fig. \ref{dipolarland}(a) can be adequately reconstructed by calculating the 2D function
$$ {\tt S}_1(\tau,t) = \beta_{1}(\tau) \opc{F}_1(t) \equiv \opc{N}_{t_1}(\tau) \opc{M}_{\tau_1}(t)
$$
However, this form of $ {\tt S}_1(\tau,t)$ does not describe the experimental signals of the other studied samples, then,  ($iii$) In order to select the contribution of the next QI,  
$\opc{F}_2(t) $, 
we analyze the difference between the measured data and the contribution of the first QI to the signal 
$$ 
\opc{C}_2(\tau,t) =\opc{M}_{\tau}(t)- {\tt S}_1(\tau,t).
$$
This data set attains its maximum value at a time which we call $t_2=\tau_2$.  Then we choose $ \opc{F}_2(t) \equiv \opc{C}_2(\tau_2, t)$ and $\beta_2(\tau) \equiv \opc{C}_2(\tau,t_2)$, the solid curve in  Fig. \ref{funbase}.  It is worth to mention that $\opc{N}_{t_1}(\tau_2) \simeq 0$,  and therefore $\tau_2$ is the time at which the the first dipolar QI has negligible contribution to the observed signal (This feature agrees with previous works \cite{dumont,interpar,bulj09}).  

%\begin{figure}[h*]
%\vspace{6 cm}
%\special{wmf:funbase.wmf x=8cm y=6cm}
%\hspace{-2 cm}
%%\vspace{-0.2 cm}
%\caption{Signals $\opc{F}_1$ (dotted), $\opc{F}_2$ (solid) and $\opc{F}_3$ (dashed) for 5CB at 302K.}
%\label{funbase}
%\end{figure} 

This election of $\opc{F}_1$ and $\opc{F}_2$ for reconstructing the NMR signal ${\tt S}_2(\tau,t)$ yields the curves in open triangles in Figs. \ref{maxabs_1} and \ref{maxabs_2} (blue triangles on the online version). It allows for an excellent reconstruction of the experimental curve in gypsum. In fact, as seen in the top of Fig.\ref{maxabs_2} both curves (solid circles and open triangles) coincide in almost all the timescale. The dipolar signal shape had already been accounted for within a restricted interval of preparation times by Dumont {\it et al.} \cite{dumont} by assuming the occurrence of two QI. Now we learn that these two QI can give a good description within an extended timescale. 

On the contrary, in LC samples, the triangle-curves only reproduce the experimental behaviour of $t_{max}(\tau)$ for times $\tau, t < 120 \mu$s in 5CB, 6CB 8CB and MBBA, that is, when the preparation times are restricted roughly to within the first half period of the dipolar coupling. 
This failure in describing $t_{max}(\tau)$ for longer times with only two quasiinvariants led us to continue with the procedure in order to find a new curve $\opc{F}_3$. Again we analyze the residue  
$$ 
\opc{C}_3(\tau,t) =\opc{M}_{\tau}(t)- {\tt S}_2(\tau,t) 
%\opc{M}_{\tau}(t)- \beta_1(\tau) \opc{F}_1(t)- \beta_2(\tau) \opc{F}_2(t)
$$
to find the time $t_3=\tau_3$ at which this data set attains its maximum amplitude and define $\opc{F}_3(t)\equiv \opc{C}_3(\tau_3,t)$ (see dashed curve in Fig.  \ref{funbase}) and  $\beta_3(\tau) \equiv \opc{C}_3(\tau,t_3)$. The times $t_1$, $t_2$  and $t_3$ corresponding to the different compounds are summarized in the following Table. %Table 1.
\begin{center}
\begin{tabular}{|c|c|c|c|c|} %
\hline 
 Compound & Temp (K)& $t_1 (\mu$s)& $t_2 (\mu$s) & $t_3 (\mu$s) \\ 
\hline
  5CB    &302 & 30   & 70    & 134   \\ 
\hline
  5CB    &297 & 26   & 60    & 106   \\ 
\hline
  6CB    &297 & 36   & 85    & 160   \\ 
\hline
  8CB    &311 & 29   & 74    & 140   \\ 
\hline 
 gypsum  &311 & 10   & 28    &  -    \\ 
\hline
  MBBA   &311 & 22   & 65    & 120   \\ 
\hline  
\end{tabular}
\vspace{0.5cm} \end{center}

\begin{figure}[h*]
\includegraphics[width=13cm,height=9cm]{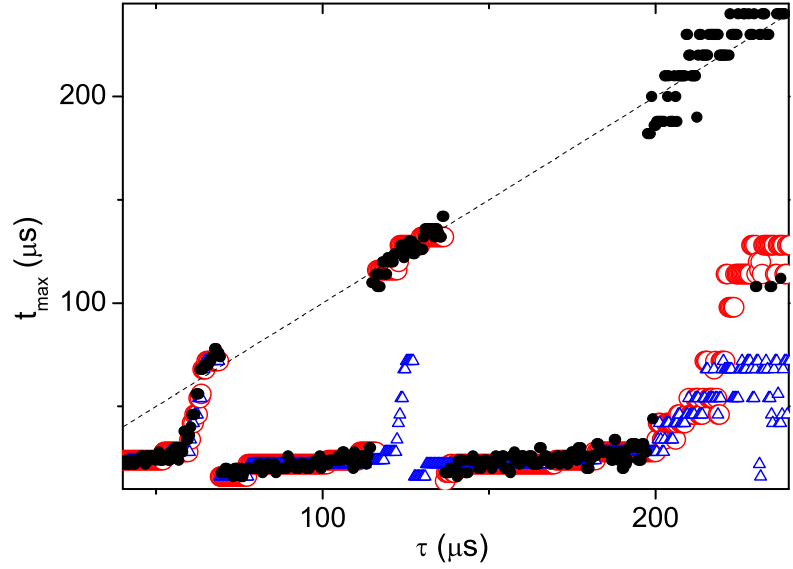}
\caption{Detail of the experimental $t_{max}(\tau)$ in nematic MBBA (solid circles) and its description with Eq.(\ref{signal_qi}) considering two QI (open triangles) and three QI (open circles).}
\label{max_MBBA}
\end{figure}

With this selection it was possible to improve the reconstructed curve for $t_{max}(\tau)$. Figures \ref{maxabs_1} (a,b,c) and \ref{maxabs_2} (b) show that involving a third quasiinvariant yields the open circles curve which gives a noticeably better reconstruction of the experimental curves. 
Fig.\ref{max_MBBA} shows a detail of $t_{max}(\tau)$ in MBBA, where the three-QI curve is definitely better than the two-QI curve especially within the interval 115 $\mu$s $< \tau < $136 $\mu$s; however it still does not account for the echo-like behaviour for longer preparation times. 

%\begin{figure}[h*]
%\vspace{7 cm}
%\special{wmf:detalleMBBA.wmf x=8cm y=6cm}
%\hspace{-2 cm}
%%\vspace{-0.2 cm}
%\caption{Detail of the experimental $t_{max}(\tau)$ in nematic MBBA (solid circles) and its description with Eq.(\ref{signal_qi}) considering two QI (open triangles) and three QI (open circles).}
%\label{max_MBBA}
%\end{figure} 

\begin{figure}[h*]
\includegraphics[width=13cm,height=9cm]{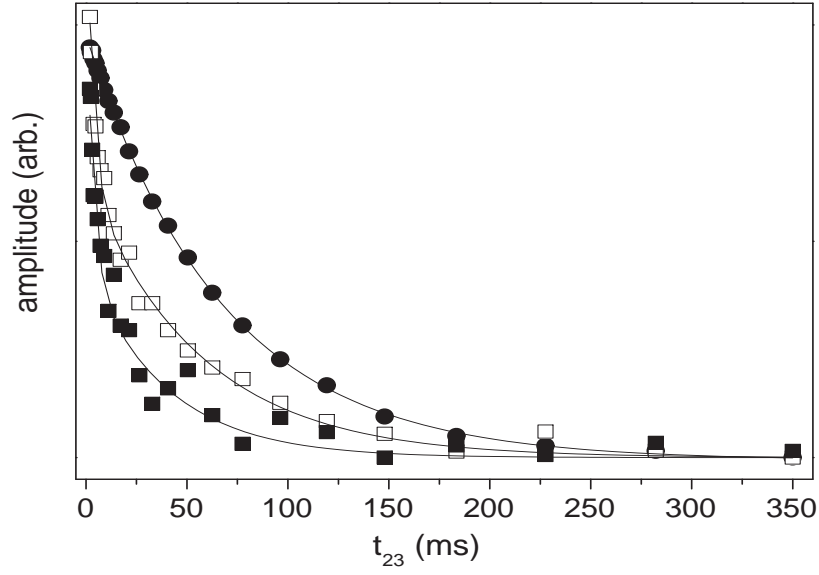}
\caption{Spin-lattice relaxation curves of three QI in 5CB at 302K prepared with different preparation times: $\tau = 30 \mu$s, circles; $\tau = 70 \mu$s, open squares and $\tau = 134 \mu$s, full squares.}
\label{T1s}
\end{figure}

In order to test if $\opc{F}_3(t)$ does in fact behave as the signal of a QI, we analyze the signal amplitude attenuation as a function of the evolution time $t_e$ as in a regular spin-lattice relaxation experiment. Figure \ref{T1s} shows amplitude attenuation of $\opc{F}_3(t)$ in 5CB at 302 K, for the preparation times shown in Table 1: $\tau = 30 \mu$s, circles; $\tau = 70 \mu$s, open squares and $\tau = 134 \mu$s, full squares. The three curves can be adequately fitted with two exponential decays, the common short time decay (the same for all) is 4 ms while the longer characteristic times are $T_{D1}$ = 67 ms (circles), $T_{D2}$ = 56 ms (open squares) and $T_{D3}$ = 39 ms (full squares).  The  fast decay is just a witness of the attenuation of higher order coherences, in fact, its characteristic time coincides with the timescale of irreversible decoherence as measured in ref. \cite{nosPRE11}. On the other hand, the longer decay times are all of the order of the {\em dipolar relaxation time} (67 ms in this case). 
%\begin{figure}[h*]
%\vspace{7 cm}
%\special{wmf:T1s.wmf x=8cm y=7cm}
%\hspace{-2 cm}
%%\vspace{-0.2 cm}
%\caption{Spin-lattice relaxation curves of three QI in 5CB at 302K prepared with different preparation times: $\tau = 30 \mu$s, circles; $\tau = 70 \mu$s, open squares and $\tau = 134 \mu$s, full squares.}
%\label{T1s}
%\end{figure} 

\begin{figure}[h*]
\includegraphics[width=13cm,height=9cm]{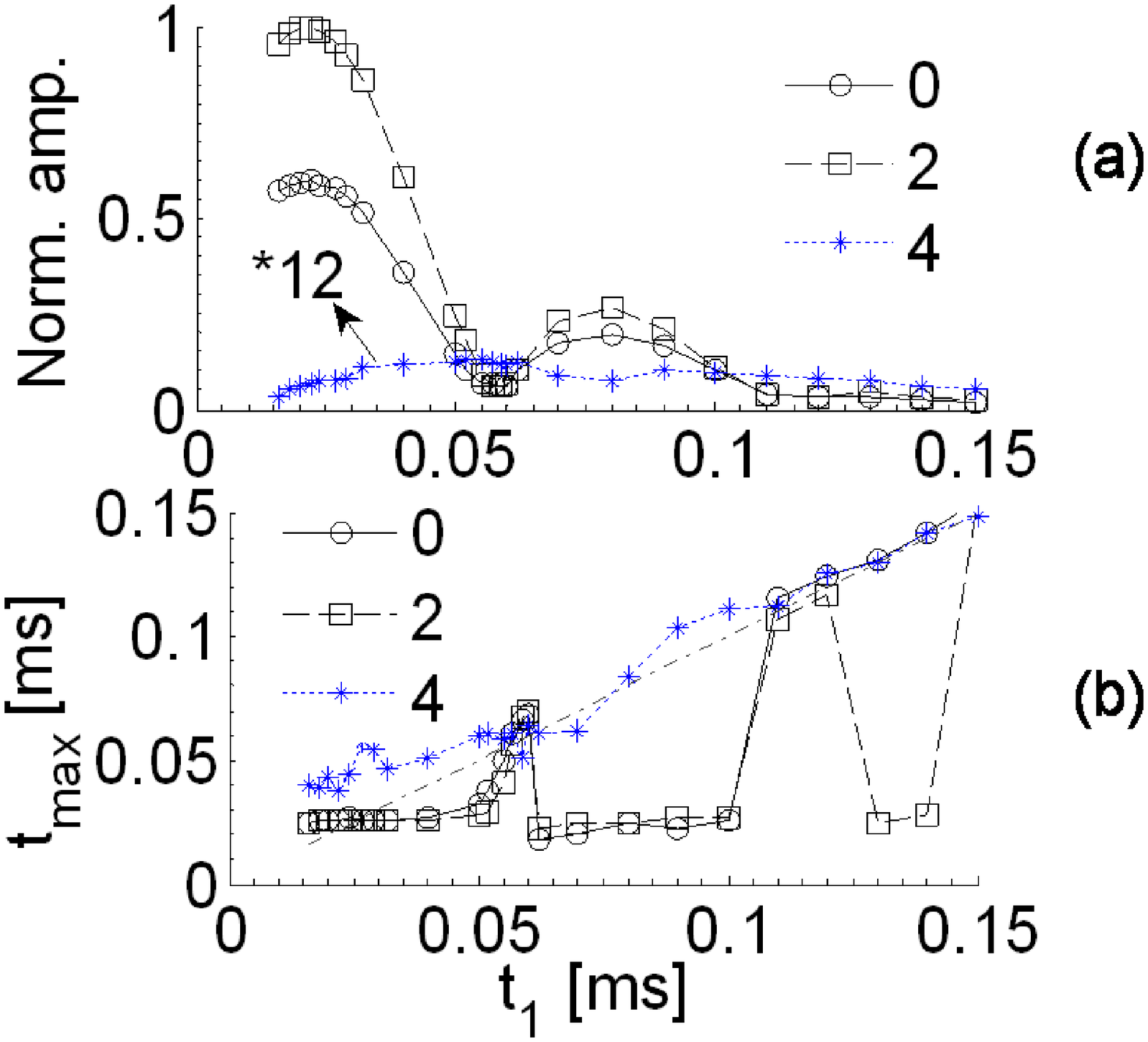}
\caption{a) Maximum amplitude attained by each coherence, b) observation-time window at which the maximum amplitudes occur.}
\label{MQC}
\end{figure}

In ref. \cite{bulj09} it was shown that the quasi equilibrium state which is proportional to $\H_{\opc{S}}$ in 5CB (prepared by setting  $\beta_{\opc{W}}(\tau_1)\simeq 0$) is a two-spin correlated state since it involves at most two-spin tensors, while states proportional to  $\H_{\opc{W}}$ (prepared by setting $\beta_{\opc{S}}(\tau_1)\simeq 0$) involve correlations of more-than-two spin tensors.
With the aim of studying the correlated nature  of the states prepared at different $\tau$'s in 5CB, we analyze the multiple quantum coherence (MQc) content on the X-basis using the same pulse sequence as in refs. \cite{bulj09,cho03}. In this experiment, rotating the state around an axis orthogonal to Z allows encoding MQc which reflect the number of multiply connected spins in the prepared states. The experiment begins by preparing the state with the phase shifted JB pulse pair, then, a waiting time $t_w$ is followed by two pulses 90º$_{(\phi+\pi/2)}$- $\epsilon$ - 90º$_{(y)}$ which encode the coherence numbers of the quantum state at time $t_w$ in the X-basis, when varying $\phi$ systematically in succesive experiments. Fourier transformation of the signal amplitude with respect to $\phi$ yields an X-basis coherence spectrum. Our experiment, conducted on 5CB at 297 K and 60 MHz, was set to encode up to 8-quantum coherences on the X-basis, however, coherences higher than 4 in 5CB were seen to fall below the noise level.  The MQc content varies with the preparation time $\tau$, as shown in Fig. \ref{MQC} (a). As expected, zero and double quantum coherences are dominant for $0< \tau < 50 \mu$s (notice that $t_i$ in 5CB change with temperature, as evident in the first two rows of Table 1). The fact that the quotient of their amplitudes is ca. 1.5 indicates that the states prepared within this interval can be described with only one QI (or dipolar constant of motion) \cite{cho03}, which is proportional to $\op{T}_{20}$, the 0 component of a spherical tensor of rank 2.  Coherences $\pm$ 4 instead, grow slowly with $\tau$ and dominate only within restricted intervals. 

%\begin{figure}[h*]
%\vspace{7 cm}
%\special{wmf:MQC.wmf x=9cm y=7cm}%maxabs
%%\hspace{-2 cm}
%\vspace{-0.2 cm}
%\caption{a) Maximum amplitude attained by each coherence, b) observation-time window at which the maximum amplitudes occur.}
%\label{MQC}
%\end{figure} 

At a preparation time near 60$\mu$s the ratio between coherences 2 and 0 changes drastically and their amplitudes become comparable to the rising 4-quantum coherence. Notice that this is precisely the preparation time at which $\opc{C}_2(\tau_2, t)$, the signal associated with the second QI, attains its maximum, showing that these states involve correlation between more-than-two spins. This characteristic indicates that these are {\em multi-spin-correlated} states. Notice that also near 110$\mu$s all the coherences attain comparable amplitudes, and this preparation time coincides with the third QI, $\opc{F}_3$.

We observed that the {\em relative} amplitudes of the various MQc components depend on which window along the acqusition time is used for calculating the MQc spectrum. This feature is shown in Fig.\ref{MQC}(b) where the time interval (along the acquisition period) at which the maximum value of each coherence component occurs is plotted as a function of the preparation time. 
The salient feature is that different coherences have a very different behaviour. The maximum contribution to coherences 0 and $\pm$2 on the X-basis come from the window near  $t_{max}=30\mu$s, except within the narrow preparation intervals which correspond to the second and third quasiinvariants. On the contrary, the maximum amplitude of coherences $\pm$4 (on the X-basis) falls  on the line $t_{max}=\tau$, in other words, it behaves like an echo. This distinct behaviour agrees with a view that multi-spin correlations arise at longer times than two-spin correlations, and also that the echo like signals prepared with $\tau > 70 \mu$s correspond to states of multi-spin nature. 

\section{Discussion and conclusions}
In this work we present an experimental survey on the occurrence of several QI created using the JB sequence, in different dipolar-coupled spin systems. To account for the dependence of the JB signals on the preparation time, we propose a generalization of the method presented in refs.\cite{dumont,interpar}.
We demonstrate that multiple QI can be prepared in the dipole interacting proton spin system on LC molecules.  The well-known dipolar order state, which is a two-spin object, comes at short preparation times, while the other QI which emerge sequentially at longer preparation times involve multiple-spin interactions. We  propose a criterion to experimentally isolate each QI and oberve that the first three states in 5CB have evidently distinct spin-lattice relaxation times which implies that they are true quasiinvariants.

Though at present the tensor structure of these quasi-invariants is still unknown, the multiple quantum coherence experiments demonstrate that the new states have multi-spin character, which implies that their tensor structure must involve products of individual angular momentum operators of many spins. Also, we verified that the observed echo-like behavior of the dipolar signal in LC for long preparation times is also a consequence of the occurrenece of many QI.  

In this way we demonstrate that the initial Zeeman order (of thermal equilibrium) can in fact be transferred to at least three dipolar constants of motion in LC (within the scanned short and intermediate $\tau$ range), while only two constants of motion were observed in gypsum when $\vec B \parallel$ [010] and only one at an orientation that makes the protons at the water molecules equivalent. 
The fact that compounds having very different numbers of interacting spins (11 to 25 in LC and infinite in gypsum) admit the preparation of more than one dipolar QI, indicates that their occurrence is not exclusively determined by the size of the dipolar network. On the contrary, both the occurrence of multiple QI and the many-spin character they show, do depend strongly  on the topology of the spin network, since it determines the quantum dynamics of the spin system. This assertion also agrees with the fact that the quality of our three QI model description varies from one LC to another. 
 
It is worth to note that the feasibility of transferring the Zeeman order to multi-spin quasi-equilibrium states is not an  exclusive feature of spin clusters as the proton system in LC's as illustrated in this work. In fact,  the ``interpair'' state observed in hydrated crystals admits multiple quantum coherences up to the fourth order on the $X-$basis, implying four-spin correlations \cite{bulj09}. 
Then the question arises about which are the necessary characteristics the dipolar network must have in order to admit the preparation of more-than-one dipolar QI.
A coarse criterion could be the feasibility of truncating the weaker term of the dipolar Hamiltonian with respect to the stronger term, which generally implies the possibility of clasifying the dipolar couplings according to their intensity into ``strong'' and ``weak''. In practice, this characteristic is also associated to the occurrence of a doublet in the NMR spectrum. 

According to our results, in LC's two-spin correlations dominate the coherent dynamics during the early timescale of the preparation period, however, there are narrow time windows where these correlations vanish and higher order correlations can efficiently give place to multi-spin QE states. Finally, for longer $\tau$ values  only the multi-spin correlations are responsible for the echo-like behavior of the NMR signal.

\newpage

\begin{thebibliography}{aeiou}
\bibitem{FID} The NMR signal ensuing a single, radio frequency pulse is called FID for Free Induction Decay signal.
\end{thebibliography}
\end{document}